\documentclass{article}

\usepackage{arxiv}

\usepackage[utf8]{inputenc}
\usepackage[T1]{fontenc}
\usepackage{hyperref}
\usepackage{url}
\usepackage{booktabs}
\usepackage{amsfonts}
\usepackage{amsmath,amssymb}
\usepackage{nicefrac}
\usepackage{microtype}
\usepackage{graphicx}
\usepackage{subcaption}
\usepackage{multirow}
\usepackage{tabularx}
\usepackage{array}
\usepackage{float}
\graphicspath{{./images/}}

\title{KEMP-PIP: A Feature-Fusion Based Approach\\for Pro-inflammatory Peptide Prediction}

\author{
  Soumik Deb Niloy$^{\dagger}$ \\
  Department of Computer Science and Engineering \\
  BRAC University \\
  Dhaka 1212, Bangladesh \\
  \texttt{soumik.deb.niloy@g.bracu.ac.bd} \\
  \And
  Md.\ Fahmid-Ul-Alam Juboraj$^{\dagger}$ \\
  Department of Computer Science and Engineering \\
  BRAC University \\
  Dhaka 1212, Bangladesh \\
  \texttt{fahmid.juboraj@bracu.ac.bd} \\
  \And
  Swakkhar Shatabda \\
  Department of Computer Science and Engineering \\
  BRAC University \\
  Dhaka 1212, Bangladesh \\
  \texttt{swakkhar.shatabda@bracu.ac.bd} \\
}

\begin{document}
\maketitle

\begin{abstract}
Pro-inflammatory peptides (PIPs) play critical roles in immune signaling and inflammation but
are difficult to identify experimentally due to costly and time-consuming assays. To address
this challenge, we present \textbf{KEMP-PIP}, a hybrid machine learning framework that
integrates deep protein embeddings with handcrafted descriptors for robust PIP prediction.
Our approach combines contextual embeddings from pretrained ESM protein language models with
multi-scale k-mer frequencies, physicochemical descriptors, and modlAMP sequence features.
Feature pruning and class-weighted logistic regression manage high dimensionality and class
imbalance, while ensemble averaging with an optimized decision threshold enhances the
sensitivity--specificity balance. Through systematic ablation studies, we demonstrate that
integrating complementary feature sets consistently improves predictive performance. On the
standard benchmark dataset, KEMP-PIP achieves an MCC of 0.505, accuracy of 0.752, and AUC
of 0.762, outperforming ProIn-fuse, MultiFeatVotPIP, and StackPIP. Relative to StackPIP,
these results represent improvements of 9.5\% in MCC and 4.8\% in both accuracy and AUC.
The KEMP-PIP web server is freely available at
\url{https://nilsparrow1920-kemp-pip.hf.space/} and the full implementation at
\url{https://github.com/S18-Niloy/KEMP-PIP}.
\end{abstract}

\noindent\small{$^\dagger$These authors contributed equally to this work.}\\[6pt]

\section{Introduction}
\label{sec:intro}

Pro-inflammatory peptides (PIPs) are short amino acid chains that promote inflammation by
activating immune responses. They function as cytokines or chemokines---signaling molecules
that recruit immune cells to sites of infection or injury, enhance pathogen clearance, and
stimulate the production of further pro-inflammatory mediators. Excessive production or
unregulated activity of PIPs, however, can lead to tissue damage, chronic inflammation, and
the pathology of numerous inflammatory and autoimmune diseases~\cite{gupta2016proinflam}.
Notable examples include fragments derived from IL-1$\beta$, which directly promote
inflammation and immune cell activation~\cite{gupta2016proinflam}.

PIPs share a characteristic set of physicochemical features. They are typically enriched in
positively charged residues (lysine, arginine) that facilitate electrostatic interactions with
negatively charged cell membranes and immune receptors. Their amphipathic
nature~\cite{PerezPaya1995}---arising from the spatial segregation of hydrophobic and
hydrophilic residues---enables membrane insertion and receptor binding, effects further
reinforced by moderate hydrophobicity~\cite{hydrophobicity} and a high hydrophobic moment.
Structural preferences for $\alpha$-helical or $\beta$-sheet conformations stabilize functional
motifs, while their compact size (10--50 residues) confers flexibility. Resistance to
proteolytic degradation prolongs their activity in inflammatory environments, and aggregation
propensity can amplify immune stimulation. Collectively, charge, hydrophobicity,
amphipathicity, secondary structure, size, and stability define the pro-inflammatory potential
of a bioactive peptide.

Experimentally, PIPs are identified by exposing immune cells or animal models to candidate
peptides and measuring downstream inflammatory responses---cytokine release, cell activation,
or tissue inflammation~\cite{mukherjee2022role}. Such assays are labor-intensive, expensive,
and low-throughput, making computational prediction an attractive complement. From a machine
learning perspective~\cite{raza2024comprehensive}, PIP prediction presents four interrelated
challenges: (i) the sequence--activity relationship is governed by subtle motifs and residue
preferences (e.g., enrichment in A, F, I, L, V) that are not obvious in raw sequence data;
(ii) validated datasets are small and class-imbalanced, with non-PIPs outnumbering PIPs,
biasing naive classifiers; (iii) immunogenicity depends on physicochemical, structural, and
evolutionary factors that must all be encoded as informative features; and (iv) models must
generalize to held-out sequences, demanding careful regularization, feature selection, and
ensemble design.

Several dedicated computational tools have addressed these challenges, including
ProInflam~\cite{gupta2016proinflam}, ProIn-Fuse~\cite{khatun2020proinfuse},
MultiFeatVotPIP~\cite{yan2024multifeatvotpip}, PIP-EL~\cite{manavalan2018pipel}, and
StackPIP~\cite{yao2025stackpip}. These methods exploit descriptors ranging from amino acid
composition and dipeptide frequencies to evolutionary profiles and motif patterns, paired
with support vector machines, random forests, and voting ensembles. Yet biologically
interpretable, therapeutically applicable predictions remain elusive. MultiFeatVotPIP, for
example, achieves competitive accuracy but relies on numerous potentially redundant
hand-crafted features that can overfit imbalanced data, applies unweighted voting that ignores
classifier-specific error profiles, and reports primarily internal validation results that
complicate generalization claims.

We address these gaps with \textbf{KEMP-PIP}, a lightweight hybrid framework that fuses
(i) contextual sequence representations from the pretrained ESM2 protein language model,
(ii) multi-scale k-mer frequency vectors encoding local motif patterns, and
(iii) physicochemical and modlAMP global descriptors. Two complementary feature sets train
separate class-weighted logistic regression classifiers whose probability outputs are
ensemble-averaged with an MCC-optimized threshold. KEMP-PIP surpasses all compared baselines
across every reported metric while remaining computationally affordable and accessible via a
public web server.

\paragraph{Contributions.}
\begin{itemize}
  \item A lightweight hybrid architecture combining pretrained protein language model
        embeddings with multi-scale handcrafted descriptors, achieving state-of-the-art PIP
        classification performance.
  \item A systematic ablation covering individual features, four classifier families, and all
        pairwise two-model ensembles with jointly tuned mixing weights and decision thresholds.
  \item A freely accessible web server accepting FASTA files, CSV files, and manual input,
        enabling the wider bioinformatics community to deploy KEMP-PIP without programming
        expertise.
\end{itemize}

\section{Data Construction}
\label{sec:data}

\paragraph{Source and preprocessing.}
The dataset originates from the MultiFeatVotPIP benchmark
repository~\cite{yan2024multifeatvotpip}. Raw sequences were preprocessed by removing
duplicates and entries containing non-standard amino acids, and all characters were normalized
to uppercase with no gaps or special characters. CD-HIT was applied at a sequence identity
threshold of 0.60 to reduce redundancy. Table~\ref{tab:dataset} summarizes the resulting
splits. The positive class (PIPs) constitutes approximately 43\% of training samples,
introducing mild class imbalance that is addressed explicitly during model training.

\begin{table}[h]
  \caption{Dataset composition after preprocessing and redundancy reduction.
           PIP: pro-inflammatory; n-PIP: non-pro-inflammatory.}
  \label{tab:dataset}
  \centering
  \begin{tabular}{cccc}
    \toprule
    \textbf{PIP (train)} & \textbf{n-PIP (train)} &
    \textbf{PIP (test)}  & \textbf{n-PIP (test)} \\
    \midrule
    1{,}245 & 1{,}627 & 171 & 171 \\
    \bottomrule
  \end{tabular}
\end{table}

\paragraph{Task formulation.}
Given a peptide sequence $s \in \mathcal{A}^*$ over the standard 20-letter amino acid alphabet
$\mathcal{A}$, the task is binary classification: predict whether $s$ exhibits
pro-inflammatory activity. We treat this as a supervised learning problem with probabilistic
output calibrated by an MCC-optimized threshold.

\section{Methodology}
\label{sec:method}

\begin{figure}[H]
  \centering
  \includegraphics[width=\linewidth]{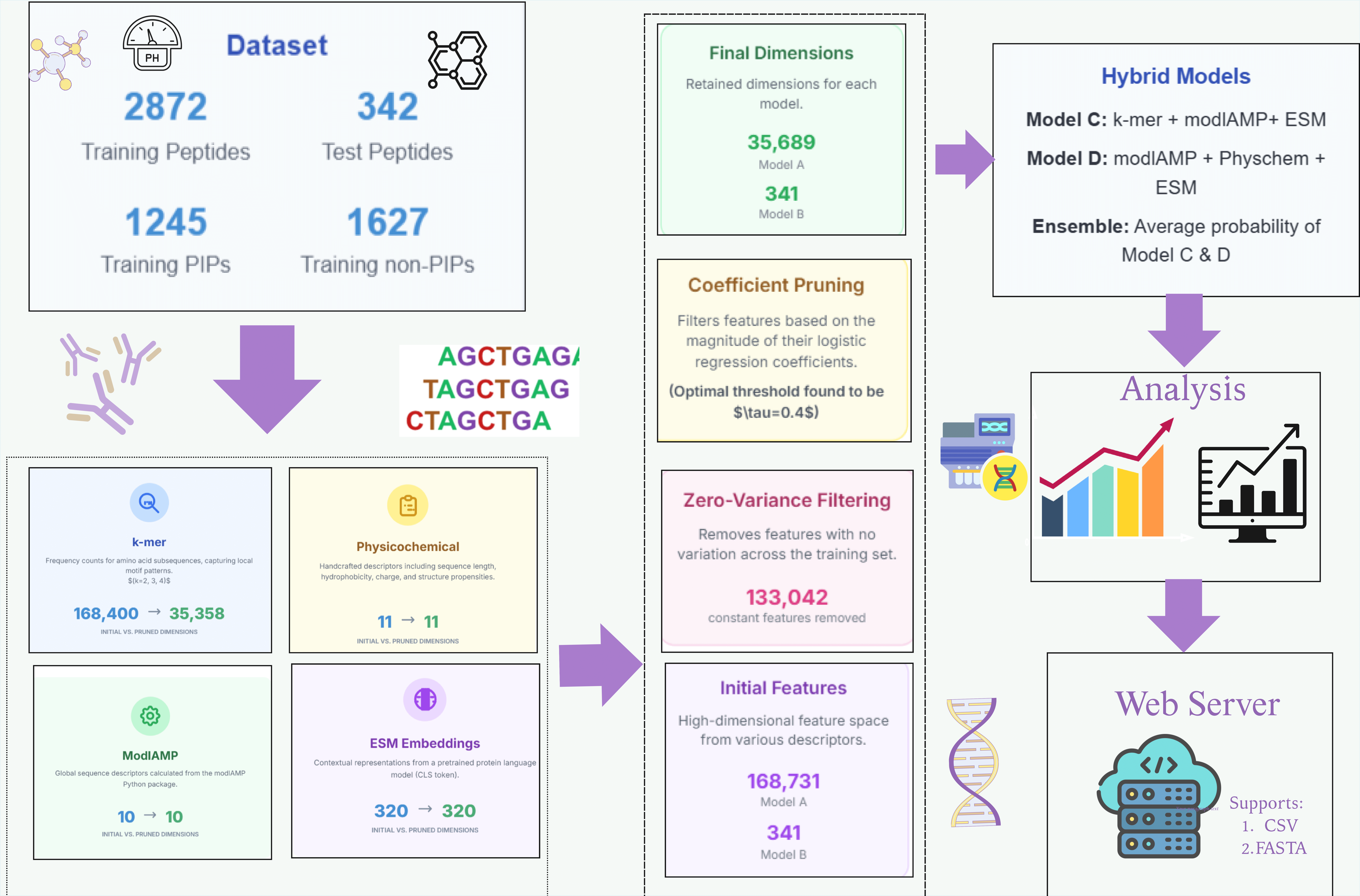}
  \caption{Architecture of the KEMP-PIP framework. Four feature streams are fused into two
           hybrid models whose predictions are ensemble-averaged with a tuned threshold.}
  \label{fig:framework}
\end{figure}

The overall workflow is illustrated in Figure~\ref{fig:framework}. Feature extraction
produces four complementary representations that are fused into two model inputs. After
zero-variance filtering and coefficient-magnitude pruning, two logistic regression classifiers
are trained and their outputs combined by probability averaging. The resulting system is
deployed as a public web server.

\subsection{Feature Extraction}

\paragraph{Multi-scale k-mer frequencies.}
For a peptide $s$ of length $L$, we compute normalized k-mer frequencies~\cite{jenike2024kmer}
for $k \in \{2, 3, 4\}$:
\begin{equation}
  f_{i,j}^{(k)} = \frac{c_{i,j}}{\sum_{m} c_{i,m}},
\end{equation}
where $c_{i,j}$ is the count of the $j$-th $k$-mer in sequence $i$. Feature dimensions are
$20^2 = 400$, $20^3 = 8{,}000$, and $20^4 = 160{,}000$, giving a concatenated vector
$\mathbf{X}_{\mathrm{kmer}} \in \mathbb{R}^{N \times 168{,}400}$.

\paragraph{Physicochemical descriptors.}
We encode 11 handcrafted descriptors per peptide: (1) sequence length $L$; (2) mean
Kyte--Doolittle hydrophobicity~\cite{Waibl2022}; (3--4) helical and sheet hydrophobic moments
(Eisenberg, $\theta = 100^\circ$ and $180^\circ$)~\cite{Eisenberg1982}; (5--6) positive
charge $Q^+$ and charge density $\rho^+$ at pH~7~\cite{shaw2010}; (7--8) helix- and
sheet-propensity fractions~\cite{Mackenzie2017}; and (9--11) trypsin, chymotrypsin, and
elastase cleavage-site densities~\cite{Sun2021}. The full physicochemical feature matrix is
$\mathbf{X}_{\mathrm{pc}} \in \mathbb{R}^{N \times 11}$.

\paragraph{ModlAMP global descriptors.}
We use the \texttt{modlAMP} Python package~\cite{mueller2017modlamp} to compute $D = 10$
global sequence descriptors per peptide, yielding
$\mathbf{X}_{\mathrm{modlAMP}} \in \mathbb{R}^{N \times 10}$.

\paragraph{ESM protein language model embeddings.}
We use the pretrained ESM2-T6-8M (UR50D) transformer~\cite{zheng2024esm}. Each peptide is
tokenized at the residue level and passed through the encoder; the \texttt{[CLS]} token
representation from the final hidden layer~\cite{SeohChangMcCallum2023} provides a fixed-length
global embedding:
\begin{equation}
  \mathbf{z}_i \in \mathbb{R}^{320}.
\end{equation}
Unlike handcrafted descriptors, these embeddings capture deep contextual dependencies learned
from millions of protein sequences.

Table~\ref{tab:feature_summary} summarizes all feature types, original dimensionalities, and
post-pruning sizes.

\begin{table}[h]
  \caption{Feature descriptor summary. Pruning applies only to k-mer features via zero-variance
           filtering; all other features are retained in full.}
  \label{tab:feature_summary}
  \centering
  \small
  \begin{tabular}{llcc}
    \toprule
    \textbf{Feature Type}    & \textbf{Description}                         &
    \textbf{Original Dim.}  & \textbf{Pruned Dim.} \\
    \midrule
    Multi-scale k-mer        & Normalized frequencies, $k \in \{2,3,4\}$    & 168,400 & 35,358 \\
    Physicochemical          & Hydrophobicity, charge, cleavage densities    & 11      & 11     \\
    ModlAMP                  & Global sequence descriptors                   & 10      & 10     \\
    ESM embeddings           & CLS token, ESM2-T6-8M                         & 320     & 320    \\
    \midrule
    \textbf{Model C (fused)} & k-mer + ModlAMP + ESM                         & 168,730 & 35,688 \\
    \textbf{Model D (fused)} & Physicochemical + ModlAMP + ESM               & 341     & 341    \\
    \bottomrule
  \end{tabular}
\end{table}

\subsection{Feature Pruning}
\label{sec:pruning}

The initial feature space of Model~C is extremely high-dimensional (168,730 variables),
necessitating a two-stage pruning strategy~\cite{YeEtAl2022_DynamicFeaturePruning}.

\paragraph{Stage 1 --- Zero-variance filtering.}
Features that are constant across all training sequences (i.e., zero variance) are discarded.
This step is critical for sparse k-mer encodings in which many $k$-mers never appear in the
corpus~\cite{xiao2023onebit}. After filtering, Model~C is reduced from 168,730 to
\textbf{35,688} features (133,042 constant features removed). Model~D retains all 341
features.

\paragraph{Stage 2 --- Coefficient-magnitude pruning.}
A class-weighted logistic regression is trained on the filtered features, and features are
ranked by $|\hat\beta_j|$~\cite{sun2024wanda}. A threshold grid
$\tau \in \{0, 10^{-6}, 5{\times}10^{-6}, \ldots, 10^{-3}\}$ is evaluated; for each $\tau$
only features with $|\hat\beta_j| \geq \tau$ are retained, a model is re-trained, and MCC
on a stratified 10\% validation split serves as the selection criterion. The optimal threshold
for both models is $\tau = 0$, confirming that retaining all non-constant features maximizes
validation MCC and that the major dimensionality reduction is achieved by Stage~1 alone.

\subsection{Feature Combination and Ensemble Learning}
\label{sec:ensemble}

Two complementary feature matrices are constructed and independently standardized:
\begin{align}
  \mathbf{X}_C &= \bigl[\mathbf{X}_{\mathrm{kmer}}
                  \mid \mathbf{X}_{\mathrm{modlAMP}}
                  \mid \mathbf{X}_{\mathrm{ESM}}\bigr]
                  \in \mathbb{R}^{N \times 35{,}688}, \label{eq:modelC}\\
  \mathbf{X}_D &= \bigl[\mathbf{X}_{\mathrm{pc}}
                  \mid \mathbf{X}_{\mathrm{modlAMP}}
                  \mid \mathbf{X}_{\mathrm{ESM}}\bigr]
                  \in \mathbb{R}^{N \times 341}. \label{eq:modelD}
\end{align}
A class-weighted logistic regression~\cite{AbboodEtAl2023} with the L-BFGS
solver~\cite{niu2023mL-BFGS} is trained on each matrix. Class weights are set inversely
proportional to class frequencies to counteract the PIP/non-PIP imbalance.

Ensemble predictions are obtained by averaging the two models' predicted probabilities
with a tunable mixing coefficient $\alpha$~\cite{yang2024gas}:
\begin{equation}
  P_{\mathrm{ens}} = \alpha\, P_C + (1-\alpha)\, P_D.
\end{equation}
The final label is assigned by thresholding:
\begin{equation}
  \hat{y} = \begin{cases} 1 & P_{\mathrm{ens}} > t, \\ 0 & \text{otherwise.} \end{cases}
\end{equation}
Both $\alpha \in [0,1]$ and $t \in (0,1)$ are jointly optimized on a held-out validation
split to maximize MCC~\cite{das2023mlwavelet}. For the primary KEMP-PIP configuration
(C$\oplus$D), the optimal values are $\alpha = 0.65$ and $t = 0.36$.

\subsection{Performance Metrics}
\label{sec:metrics}

We evaluate all models using five standard bioinformatics classification
metrics~\cite{Fu2024}: Accuracy (ACC), Sensitivity (SN), Specificity (SP), Matthews
Correlation Coefficient (MCC)~\cite{xiao2024pelpvp}, and Area Under the ROC Curve
(AUC)~\cite{sk2023}:
\begin{align}
  \mathrm{ACC} &= \frac{\mathrm{TP}+\mathrm{TN}}
                       {\mathrm{TP}+\mathrm{TN}+\mathrm{FP}+\mathrm{FN}},
  &
  \mathrm{SN}  &= \frac{\mathrm{TP}}{\mathrm{TP}+\mathrm{FN}},
  &
  \mathrm{SP}  &= \frac{\mathrm{TN}}{\mathrm{TN}+\mathrm{FP}}, \\[6pt]
  \mathrm{MCC} &= \frac{\mathrm{TP}{\cdot}\mathrm{TN}
                        -\mathrm{FP}{\cdot}\mathrm{FN}}
                       {\sqrt{(\mathrm{TP}{+}\mathrm{FP})
                              (\mathrm{TP}{+}\mathrm{FN})
                              (\mathrm{TN}{+}\mathrm{FP})
                              (\mathrm{TN}{+}\mathrm{FN})}}.
\end{align}
MCC is the primary selection criterion throughout because it provides a balanced summary of
classification performance under class imbalance, penalizing both false positives and false
negatives.

\section{Results and Analysis}
\label{sec:results}

\subsection{Feature Ablation}
\label{sec:feat_ablation}

Table~\ref{tab:ablation} reports MCC, AUC, and ACC on both the validation split and the
independent test set for each individual feature type and multi-feature combination. The
best value in each column is bolded.

\begin{table}[h]
  \caption{Feature ablation study. Best values per column in \textbf{bold}.}
  \label{tab:ablation}
  \centering
  \small
  \setlength{\tabcolsep}{4pt}
  \begin{tabular}{lr ccc ccc}
    \toprule
    \multirow{2}{*}{\textbf{Feature Combination}} &
    \multirow{2}{*}{\textbf{Dim.}} &
    \multicolumn{3}{c}{\textbf{Validation}} &
    \multicolumn{3}{c}{\textbf{Test}} \\
    \cmidrule(lr){3-5}\cmidrule(lr){6-8}
    & & MCC & AUC & ACC & MCC & AUC & ACC \\
    \midrule
    \multicolumn{8}{l}{\textit{Individual feature sets}} \\
    k-mer              & 35,358 & 0.2391 & 0.6868 & 0.6319 & 0.3351 & 0.7592 & 0.6608 \\
    Physicochemical    & 11     & 0.0879 & 0.6360 & 0.5035 & 0.1451 & 0.6438 & 0.5643 \\
    ESM                & 320    & 0.1797 & 0.6009 & 0.5660 & 0.2026 & 0.6577 & 0.5994 \\
    modlAMP            & 10     & 0.1658 & 0.6432 & 0.5382 & 0.1461 & 0.6538 & 0.5643 \\
    \midrule
    \multicolumn{8}{l}{\textit{Two-feature combinations}} \\
    k-mer + Physchem   & 35,369 & 0.2401 & 0.6912 & 0.6319 & 0.3704 & 0.7682 & 0.6784 \\
    k-mer + ESM        & 35,678 & 0.2807 & 0.6997 & 0.6493 & 0.3486 & 0.7755 & 0.6667 \\
    k-mer + modlAMP    & 35,368 & 0.2546 & 0.6921 & 0.6389 & 0.3735 & 0.7739 & 0.6813 \\
    \midrule
    \multicolumn{8}{l}{\textit{Three- and four-feature combinations}} \\
    k-mer + Physchem + ESM
      & 35,689 & \textbf{0.2873} & 0.7034 & \textbf{0.6528}
               & 0.3771          & 0.7809          & 0.6813          \\
    k-mer + ESM + modlAMP
      & 35,688 & 0.2818          & \textbf{0.7047} & 0.6493
               & 0.3704          & \textbf{0.7834} & 0.6784          \\
    k-mer + Physchem + ESM + modlAMP
      & 35,699 & 0.2807          & 0.7070          & 0.6493
               & \textbf{0.4042} & 0.7876          & \textbf{0.6959} \\
    \bottomrule
  \end{tabular}
\end{table}

Several trends are apparent. First, k-mer features carry the strongest standalone signal,
reflecting the importance of local sequence composition for proinflammatory activity. Second,
every additional descriptor---physicochemical, ESM, or modlAMP---provides a consistent
incremental gain, confirming that the four feature types are complementary rather than
redundant. Third, the best validation configuration (\texttt{kmer+physchem+ESM}, MCC~0.2873)
differs from the best test configuration (\texttt{kmer+physchem+ESM+modlAMP}, MCC~0.4042),
suggesting that the validation split is conservative and that the full four-feature fusion
generalizes most robustly to unseen data.

\subsection{Model Ablation Across Classifiers}
\label{sec:model_ablation}

Table~\ref{tab:ablation_subcolumns} evaluates the same feature combinations using four
alternative classifiers---XGBoost, Random Forest (RF), Support Vector Machine (SVM), and
Multi-Layer Perceptron (MLP)---to assess whether the feature-fusion gains are
classifier-agnostic.

\begin{table}[h]
  \caption{Feature ablation across four classifiers on the independent test set.
           Best MCC per row in \textbf{bold}.
           Abbreviations: k = k-mer, p = physicochemical, m = modlAMP, e = ESM.}
  \label{tab:ablation_subcolumns}
  \centering
  \scriptsize
  \setlength{\tabcolsep}{3pt}
  \begin{tabular}{l ccc ccc ccc ccc}
    \toprule
    \multirow{2}{*}{\textbf{Feature Set}} &
    \multicolumn{3}{c}{\textbf{XGBoost}} &
    \multicolumn{3}{c}{\textbf{RF}} &
    \multicolumn{3}{c}{\textbf{SVM}} &
    \multicolumn{3}{c}{\textbf{MLP}} \\
    \cmidrule(lr){2-4}\cmidrule(lr){5-7}\cmidrule(lr){8-10}\cmidrule(lr){11-13}
    & MCC & AUC & ACC & MCC & AUC & ACC & MCC & AUC & ACC & MCC & AUC & ACC \\
    \midrule
    k              & 0.2761 & 0.6973 & 0.6374 & 0.2458 & 0.7325 & 0.6228 & $-$0.0524 & 0.3909 & 0.4825 & \textbf{0.3107} & 0.7575 & 0.6491 \\
    p              & 0.1530 & 0.6202 & 0.5760 & \textbf{0.1588} & 0.6487 & 0.5789 & 0.1230 & 0.5990 & 0.5614 & 0.0946 & 0.6266 & 0.5468 \\
    m              & \textbf{0.3275} & 0.7093 & 0.6637 & 0.2807 & 0.7061 & 0.6404 & 0.2515 & 0.6464 & 0.6257 & 0.2111 & 0.6706 & 0.6023 \\
    e              & 0.1463 & 0.6141 & 0.5731 & 0.1859 & 0.6239 & 0.5906 & 0.2132 & 0.6433 & 0.6053 & \textbf{0.2281} & 0.6417 & 0.6140 \\
    k+p            & 0.3047 & 0.7244 & 0.6520 & 0.2411 & 0.7033 & 0.6199 & $-$0.0432 & 0.3931 & 0.4854 & 0.2414 & 0.7222 & 0.5994 \\
    k+m            & 0.2870 & 0.7134 & 0.6433 & 0.2781 & 0.7194 & 0.6374 & $-$0.0432 & 0.3940 & 0.4854 & \textbf{0.4219} & 0.7561 & 0.7105 \\
    k+e            & 0.1696 & 0.6267 & 0.5848 & 0.1460 & 0.6267 & 0.5702 & $-$0.0778 & 0.4046 & 0.4737 & \textbf{0.3100} & 0.7399 & 0.6550 \\
    p+m            & 0.1872 & 0.6586 & 0.5936 & 0.2515 & 0.6940 & 0.6257 & 0.1758 & 0.6202 & 0.5877 & 0.1770 & 0.6420 & 0.5877 \\
    p+e            & \textbf{0.2460} & 0.6638 & 0.6228 & 0.1845 & 0.6532 & 0.5906 & 0.2003 & 0.6473 & 0.5994 & 0.1943 & 0.6258 & 0.5965 \\
    m+e            & 0.2176 & 0.6570 & 0.6082 & 0.1763 & 0.6569 & 0.5877 & 0.2007 & 0.6488 & 0.5994 & \textbf{0.2311} & 0.6487 & 0.6082 \\
    k+p+m          & 0.2515 & 0.7159 & 0.6257 & 0.2532 & 0.7009 & 0.6257 & $-$0.0425 & 0.3962 & 0.4854 & \textbf{0.2762} & 0.7543 & 0.6316 \\
    k+p+e          & 0.2466 & 0.6815 & 0.6228 & 0.2105 & 0.6489 & 0.6023 & $-$0.0699 & 0.4061 & 0.4766 & 0.1410 & 0.7535 & 0.5380 \\
    k+m+e          & 0.2526 & 0.6798 & 0.6257 & 0.1553 & 0.6585 & 0.5760 & $-$0.0605 & 0.4072 & 0.4795 & 0.1669 & 0.7423 & 0.5409 \\
    p+m+e          & 0.2404 & 0.6562 & 0.6199 & 0.2663 & 0.6665 & 0.6316 & 0.2121 & 0.6511 & 0.6053 & \textbf{0.2285} & 0.6512 & 0.6140 \\
    k+p+m+e        & 0.2281 & 0.6703 & 0.6140 & 0.1974 & 0.6540 & 0.5965 & $-$0.0686 & 0.4092 & 0.4766 & \textbf{0.2713} & 0.7192 & 0.6199 \\
    \bottomrule
  \end{tabular}
\end{table}

The SVM consistently underperforms with k-mer features due to the high dimensionality of the
input, often yielding negative MCC values---a known failure mode of kernel SVMs on
ultra-sparse, high-dimensional data without specialized preprocessing. XGBoost and the MLP
show the greatest benefit from feature fusion, with the MLP achieving the highest single-model
MCC of 0.4219 on \texttt{k+m}. Across classifiers, modlAMP exhibits surprisingly strong
standalone performance (XGBoost MCC~0.3275), highlighting the value of global sequence
descriptors even without deep embeddings.

\subsection{Ensemble Ablation}
\label{sec:ensemble_ablation}

To identify the optimal two-model combination, we trained five single logistic regression
models on the following feature sets:
\begin{itemize}
  \item \texttt{kmer + physchem + ESM}
  \item \texttt{kmer + physchem + modlAMP}
  \item \texttt{kmer + ESM + modlAMP}
  \item \texttt{physchem + ESM + modlAMP}
  \item \texttt{kmer + physchem + ESM + modlAMP}
\end{itemize}
All ten pairwise ensembles were formed by linearly blending predicted probabilities with
weight $\alpha \in [0,1]$, and both $\alpha$ and the decision threshold $t$ were jointly
optimized to maximize MCC on the held-out validation split.

\begin{table}[h]
  \caption{Performance of pairwise ensembles with optimized $\alpha$ and threshold $t$.
           Best value per column in \textbf{bold}. $\oplus$ denotes ensemble averaging.}
  \label{tab:ensemble_ablation}
  \centering
  \small
  \setlength{\tabcolsep}{5pt}
  \begin{tabular}{lcc cccccc}
    \toprule
    \textbf{Ensemble} & $\boldsymbol{\alpha}$ & $\boldsymbol{t}$ &
    \textbf{ACC} & \textbf{AUC} & \textbf{MCC} & \textbf{F1} & \textbf{SN} & \textbf{SP} \\
    \midrule
    A $\oplus$ B & 0.10 & 0.10 & 0.7339 & 0.7787 & 0.4738 & 0.7534 & 0.8129 & 0.6550 \\
    A $\oplus$ C & 0.00 & 0.17 & 0.7251 & \textbf{0.7834} & 0.4508 & 0.7314 & 0.7485 & 0.7018 \\
    A $\oplus$ D & 0.60 & 0.37 & 0.7456 & 0.7585 & 0.4922 & 0.7372 & 0.7135 & 0.7778 \\
    A $\oplus$ E & 0.00 & 0.18 & 0.7281 & 0.7876 & 0.4565 & 0.7335 & 0.7485 & 0.7076 \\
    B $\oplus$ C & 0.55 & 0.11 & 0.7368 & 0.7820 & 0.4784 & 0.7541 & 0.8070 & 0.6667 \\
    B $\oplus$ D & 0.95 & 0.10 & 0.7398 & 0.7750 & 0.4912 & \textbf{0.7652} & \textbf{0.8480} & 0.6316 \\
    B $\oplus$ E & 0.90 & 0.10 & 0.7368 & 0.7792 & 0.4793 & 0.7554 & 0.8129 & 0.6608 \\
    \textbf{C $\oplus$ D} & \textbf{0.65} & \textbf{0.36} & \textbf{0.7515} & 0.7617 & \textbf{0.5049} & 0.7401 & 0.7076 & \textbf{0.7953} \\
    C $\oplus$ E & 0.15 & 0.17 & 0.7281 & 0.7873 & 0.4568 & 0.7350 & 0.7544 & 0.7018 \\
    D $\oplus$ E & 0.40 & 0.37 & 0.7485 & 0.7620 & 0.4979 & 0.7410 & 0.7193 & 0.7778 \\
    \bottomrule
  \end{tabular}
\end{table}

\begin{figure}[H]
  \centering
  \renewcommand{\thesubfigure}{\alph{subfigure}}
  \setcounter{subfigure}{6} 
  
  \begin{subfigure}{0.5\linewidth}
    \includegraphics[width=\linewidth]{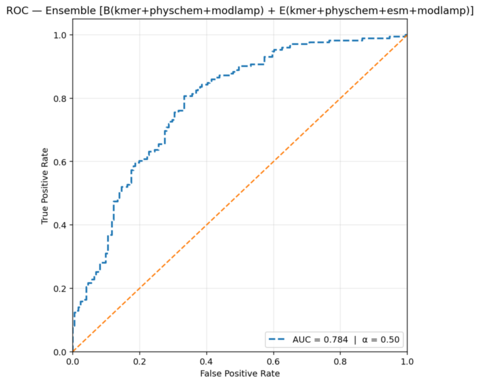}
    \caption{B $\oplus$ E}
  \end{subfigure}\hfill
  \begin{subfigure}{0.43\linewidth}
    \includegraphics[width=\linewidth]{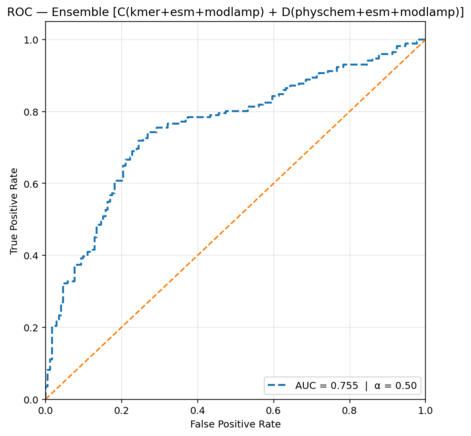}
    \caption{C $\oplus$ D}
  \end{subfigure}\\[4pt]
  \begin{subfigure}{0.48\linewidth}
    \includegraphics[width=\linewidth]{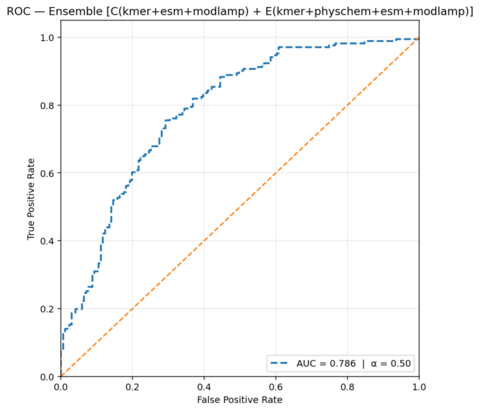}
    \caption{C $\oplus$ E}
  \end{subfigure}\hfill
  \begin{subfigure}{0.48\linewidth}
    \includegraphics[width=\linewidth]{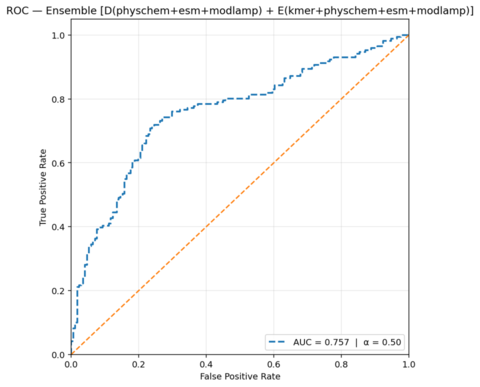}
    \caption{D $\oplus$ E}
  \end{subfigure}
  
  \renewcommand{\thesubfigure}{\alph{subfigure}} 
  \setcounter{subfigure}{0}
  
  \caption{ROC curves for ensemble combinations (a)--(j).}
  \label{fig:roc1}
\end{figure}

Table~\ref{tab:ensemble_ablation} shows that no single ensemble dominates on all metrics,
reflecting the well-known trade-off between sensitivity and specificity. The
\textbf{C$\oplus$D} ensemble achieves the best overall balance, leading on ACC (0.7515), MCC
(0.5049), and SP (0.7953); it is therefore selected as the primary KEMP-PIP configuration.
The A$\oplus$C ensemble achieves the highest AUC (0.7834), making it preferable when
ranking-based evaluation is paramount. The B$\oplus$D ensemble leads on F1 (0.7652) and SN
(0.8480), making it the most appropriate choice when minimizing false negatives is the
priority (e.g., clinical screening). ROC curves for all ten pairwise ensembles are displayed
in Figure~\ref{fig:roc1}.

\begin{figure}[H]
  \centering
  \begin{subfigure}{0.48\linewidth}
    \includegraphics[width=\linewidth]{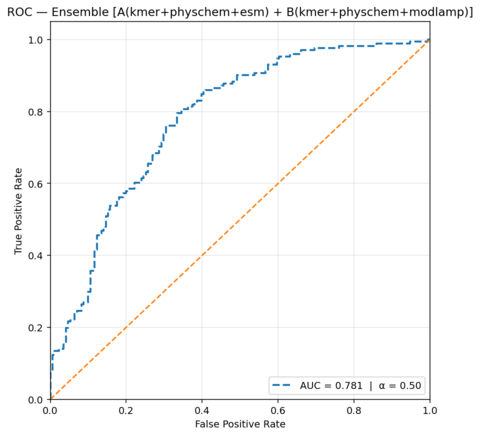}
    \caption{A $\oplus$ B}
  \end{subfigure}\hfill
  \begin{subfigure}{0.48\linewidth}
    \includegraphics[width=\linewidth]{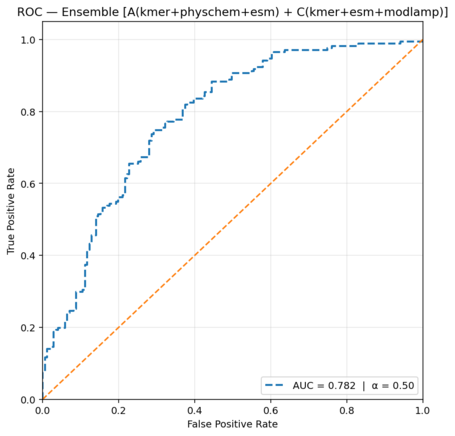}
    \caption{A $\oplus$ C}
  \end{subfigure}\\[4pt]
  \begin{subfigure}{0.48\linewidth}
    \includegraphics[width=\linewidth]{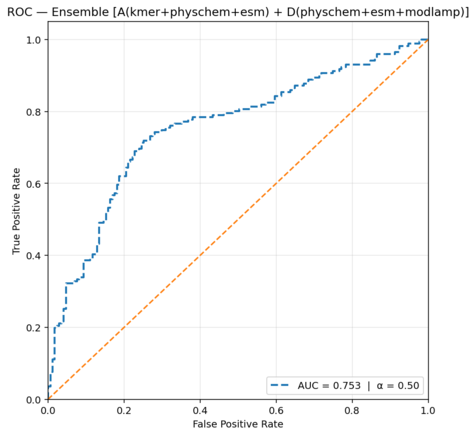}
    \caption{A $\oplus$ D}
  \end{subfigure}\hfill
  \begin{subfigure}{0.48\linewidth}
    \includegraphics[width=\linewidth]{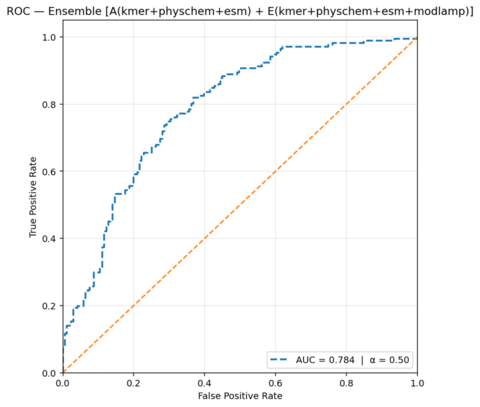}
    \caption{A $\oplus$ E}
  \end{subfigure}\\[4pt]
  \begin{subfigure}{0.48\linewidth}
    \includegraphics[width=\linewidth]{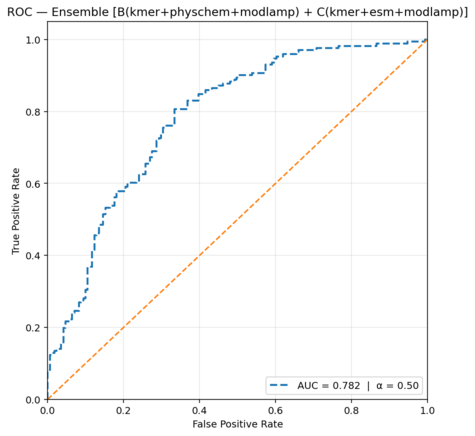}
    \caption{B $\oplus$ C}
  \end{subfigure}\hfill
  \begin{subfigure}{0.48\linewidth}
    \includegraphics[width=\linewidth]{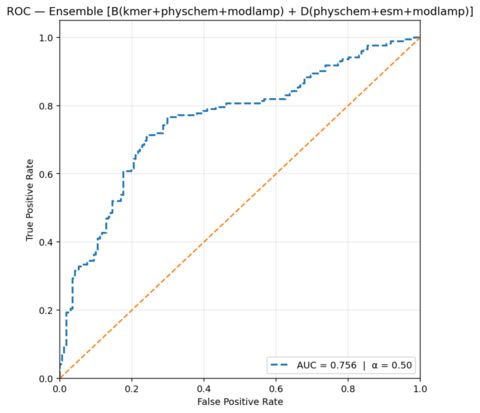}
    \caption{B $\oplus$ D}
  \end{subfigure}\\[4pt]
\end{figure}

\subsection{Comparison with State-of-the-Art Methods}
\label{sec:sota}

Table~\ref{tab:results} compares KEMP-PIP (C$\oplus$D configuration) against ProIn-fuse,
MultiFeatVotPIP, and StackPIP on the independent test set. KEMP-PIP achieves the highest
value on every reported metric.

\begin{table}[h]
  \caption{Independent test-set comparison. Best scores in \textbf{bold}.}
  \label{tab:results}
  \centering
  \begin{tabular}{lccccc}
    \toprule
    \textbf{Method}   & \textbf{SN} & \textbf{SP} & \textbf{ACC} & \textbf{AUC} & \textbf{MCC} \\
    \midrule
    ProIn-fuse        & 0.502 & 0.736 & 0.619 & 0.704 & 0.246 \\
    MultiFeatVotPIP   & 0.521 & 0.790 & 0.655 & 0.686 & 0.322 \\
    StackPIP          & 0.628 & 0.731 & 0.704 & 0.714 & 0.410 \\
    \textbf{KEMP-PIP} & \textbf{0.708} & \textbf{0.795} & \textbf{0.752} & \textbf{0.762} & \textbf{0.505} \\
    \bottomrule
  \end{tabular}
\end{table}

Relative to the previous best method (StackPIP), KEMP-PIP improves MCC by 9.5\% and both
accuracy and AUC by 4.8\%. It also improves sensitivity by 12.7 percentage points while
simultaneously improving specificity by 8.8 percentage points---a combination rarely achieved,
since sensitivity and specificity are typically in tension. This balanced gain reflects the
effectiveness of the ensemble design and MCC-based threshold optimization.

\begin{figure}[H]
  \centering
  \includegraphics[width=0.9\linewidth]{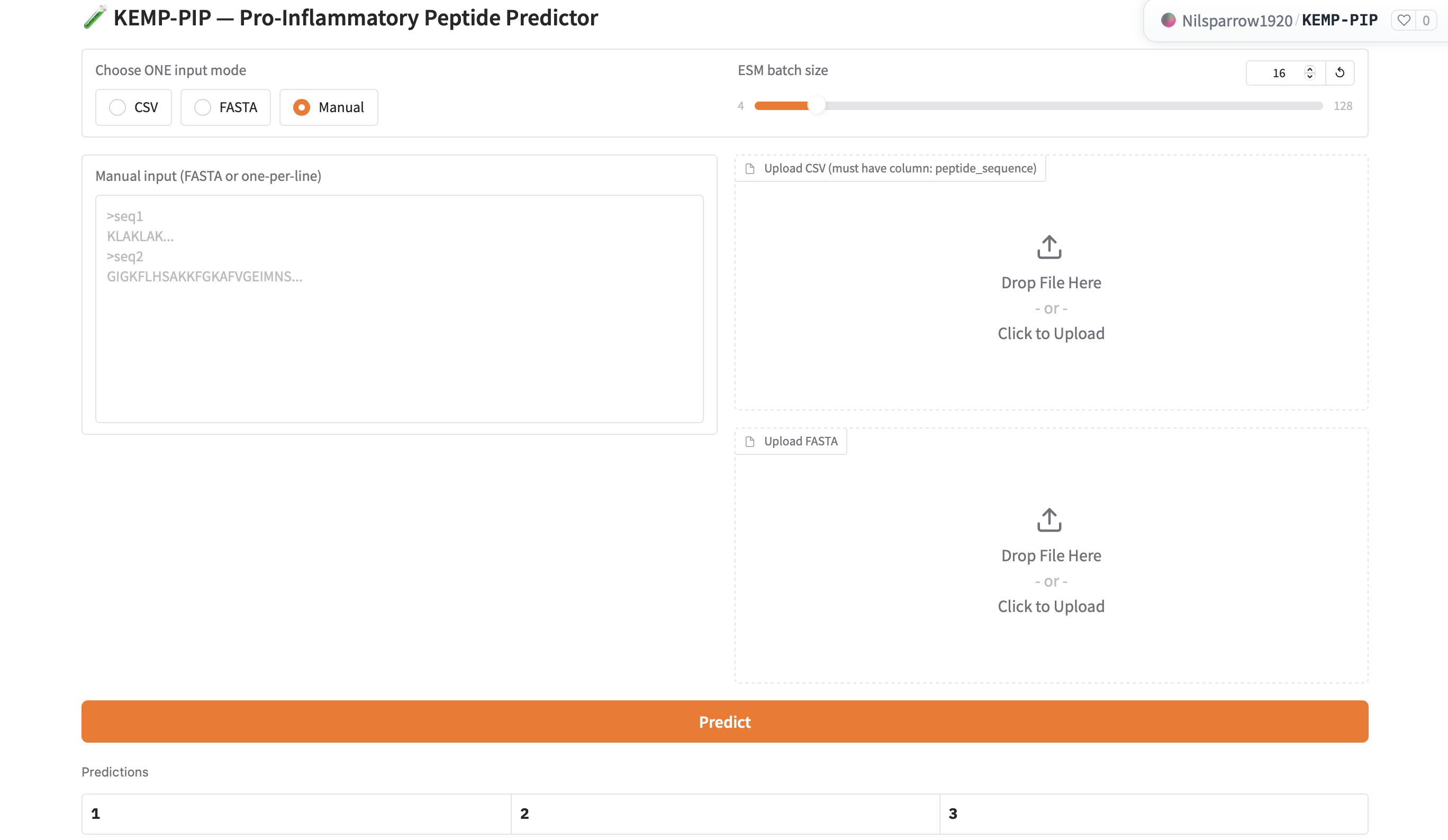}
  \caption{Web interface of the KEMP-PIP framework, supporting FASTA file upload, CSV upload,
           and manual sequence entry.}
  \label{fig:web}
\end{figure}

\paragraph{Web server.}
To maximize accessibility, we developed a user-friendly web server
(Figure~\ref{fig:web}) accepting peptide sequences in FASTA or plain text format and
returning per-sequence probability scores and binary classification labels. The server requires
no programming expertise and is freely available at
\url{https://nilsparrow1920-kemp-pip.hf.space/}.

\section{Conclusion}
\label{sec:conclusion}

We presented KEMP-PIP, a feature-fusion machine learning framework for the binary
classification of pro-inflammatory peptides. By fusing contextual ESM transformer embeddings
with multi-scale k-mer frequencies, physicochemical indices, and modlAMP global descriptors,
the framework captures complementary sequence signals---from fine-grained local motifs to
global biophysical properties and learned evolutionary context---that individually fail to
characterize pro-inflammatory activity fully. Zero-variance filtering and coefficient-based
pruning reduce the feature space efficiently, class-weighted logistic regression counteracts
label imbalance, and jointly optimized ensemble averaging balances sensitivity and specificity
without sacrificing either. KEMP-PIP surpasses ProIn-fuse, MultiFeatVotPIP, and StackPIP
across all five evaluated metrics on the independent test set.

Limitations remain. The training corpus, though curated and de-redundified, may not fully
represent the sequence and structural diversity of PIPs across organisms. Future directions
include expanding the dataset to cover taxonomically diverse peptides, incorporating
three-dimensional structural descriptors from predicted protein structures, and exploring
larger pretrained language models (e.g., ESM2-3B) or fine-tuning strategies tailored to
immunological data. Integrating KEMP-PIP into automated peptide design pipelines could
further accelerate the discovery of therapeutic candidates.

\paragraph{Author contributions.}
Soumik Deb Niloy: Methodology, Software, Investigation, Writing -- Original Draft.
Md.\ Fahmid-Ul-Alam Juboraj: Methodology, Formal Analysis, Investigation, Writing -- Original Draft.
Swakkhar Shatabda: Conceptualization, Supervision, Writing -- Review \& Editing.

\paragraph{Data and code availability.}
The dataset is publicly available at
\url{https://github.com/ChaoruiYan019/MultiFeatVotPIP}.
Source code and web server: \url{https://github.com/S18-Niloy/KEMP-PIP} and
\url{https://nilsparrow1920-kemp-pip.hf.space/}.

\bibliographystyle{unsrt}

\begin{thebibliography}{30}

\bibitem{gupta2016proinflam}
S.~Gupta, P.~Kapoor, K.~Chaudhary, A.~Gautam, R.~Kumar, Open Source Drug Discovery
Consortium, and G.~P.~S.~Raghava.
\newblock {ProInflam}: A webserver for the prediction of proinflammatory antigenicity of
peptides and proteins.
\newblock {\em Journal of Translational Medicine}, 14:1--12, 2016.

\bibitem{PerezPaya1995}
E.~P\'erez-Pay\'a, R.~A.~Houghten, and S.~E.~Blondelle.
\newblock The role of amphipathicity in the folding, self-association and biological activity
of multiple subunit small proteins.
\newblock {\em Journal of Biological Chemistry}, 270(3):1048--1056, 1995.

\bibitem{hydrophobicity}
C.~Zhu, Y.~Gao, H.~Li, S.~Meng, and L.~Li.
\newblock Characterizing hydrophobicity of amino acid side chains in a protein environment via
measuring contact angle of a water nanodroplet on planar peptide network.
\newblock {\em Proceedings of the National Academy of Sciences}, 2015.

\bibitem{mukherjee2022role}
P.~Mukherjee, S.~Roy, D.~Ghosh, and S.~K.~Nandi.
\newblock Role of animal models in biomedical research: a review.
\newblock {\em Laboratory Animal Research}, 38:18, 2022.

\bibitem{raza2024comprehensive}
A.~Raza, J.~Uddin, S.~Akbar, F.~K.~Alarfaj, Q.~Zou, and A.~Ahmad.
\newblock Comprehensive analysis of computational methods for predicting anti-inflammatory
peptides.
\newblock {\em Archives of Computational Methods in Engineering}, 31:3211--3229, 2024.

\bibitem{khatun2020proinfuse}
M.~Khatun, M.~M.~Hasan, W.~Shoombuatong, and H.~Kurata.
\newblock {ProIn-Fuse}: Improved and robust prediction of proinflammatory peptides by fusing
of multiple feature representations.
\newblock {\em Journal of Computer-Aided Molecular Design}, 34(12):1229--1236, 2020.

\bibitem{yan2024multifeatvotpip}
C.~Yan, A.~Geng, Z.~Pan, Z.~Zhang, and F.~Cui.
\newblock {MultiFeatVotPIP}: A voting-based ensemble learning framework for predicting
proinflammatory peptides.
\newblock {\em Briefings in Bioinformatics}, 25(6):bbae505, 2024.

\bibitem{manavalan2018pipel}
B.~Manavalan, T.~H.~Shin, M.~O.~Kim, and G.~Lee.
\newblock {PIP-EL}: A new ensemble learning method for improved proinflammatory peptide
predictions.
\newblock {\em Frontiers in Immunology}, 9:1783, 2018.

\bibitem{yao2025stackpip}
L.~Yao, F.~Wang, P.~Xie, J.~Guan, Z.~Zhao, X.~He, X.~Liu, Y.-C.~Chiang, and T.-Y.~Lee.
\newblock {StackPIP}: An effective computational framework for accurate and balanced
identification of proinflammatory peptides.
\newblock {\em Journal of Chemical Information and Modeling}, 65(14):7777--7788, 2025.

\bibitem{jenike2024kmer}
K.~M.~Jenike, L.~Campos-Dom\'inguez, M.~Bodd\'e, J.~Cerca, C.~N.~Hodson, M.~C.~Schatz,
and K.~S.~Jaron.
\newblock Guide to k-mer approaches for genomics across the tree of life.
\newblock arXiv:2404.01519, 2024.

\bibitem{Waibl2022}
F.~Waibl, M.~L.~Fern\'andez-Quintero, F.~S.~Wedl, et~al.
\newblock Comparison of hydrophobicity scales for predicting biophysical properties of
antibodies.
\newblock {\em Frontiers in Molecular Biosciences}, 9:960194, 2022.

\bibitem{Eisenberg1982}
D.~Eisenberg, R.~M.~Weiss, T.~C.~Terwilliger, and W.~Wilcox.
\newblock Hydrophobic moments and protein structure.
\newblock {\em Faraday Discussions of the Chemical Society}, 17:109--120, 1982.

\bibitem{shaw2010}
B.~F.~Shaw, H.~Arthanari, M.~Narovlyansky, et~al.
\newblock Neutralizing positive charges at the surface of a protein lowers its rate of amide
hydrogen exchange without altering its structure or increasing its thermostability.
\newblock {\em Journal of the American Chemical Society}, 132(49):17411--17425, 2010.

\bibitem{Mackenzie2017}
C.~O.~Mackenzie and G.~Grigoryan.
\newblock Protein structural motifs in prediction and design.
\newblock {\em Current Opinion in Structural Biology}, 44:161--167, 2017.

\bibitem{Sun2021}
H.~Sun, B.~Qiao, W.~Choi, et~al.
\newblock Origin of proteolytic stability of peptide-brush polymers as globular
proteomimetics.
\newblock {\em ACS Central Science}, 7(12):2063--2072, 2021.

\bibitem{mueller2017modlamp}
A.~T.~M\"uller, G.~Gabernet, J.~A.~Hiss, and G.~Schneider.
\newblock {modlAMP}: Python for antimicrobial peptides.
\newblock {\em Bioinformatics}, 33(17):2753--2755, 2017.

\bibitem{zheng2024esm}
K.~Zheng, S.~Long, T.~Lu, J.~Yang, X.~Dai, M.~Zhang, Z.~Nie, W.-Y.~Ma, and H.~Zhou.
\newblock {ESM} all-atom: Multi-scale protein language model for unified molecular modeling.
\newblock arXiv:2403.12995, 2024.

\bibitem{SeohChangMcCallum2023}
R.~Seoh, H.-S.~Chang, and A.~McCallum.
\newblock Encoding multi-domain scientific papers by ensembling multiple {CLS} tokens.
\newblock arXiv:2309.04333, 2023.

\bibitem{YeEtAl2022_DynamicFeaturePruning}
Y.~Ye, H.~Zhou, J.~Cai, et~al.
\newblock Dynamic feature pruning and consolidation for occluded person re-identification.
\newblock arXiv:2211.14742, 2022.

\bibitem{xiao2023onebit}
Y.-H.~Xiao, L.~Huang, D.~Ram\'irez, C.~Qian, and H.~C.~So.
\newblock One-bit covariance reconstruction with non-zero thresholds.
\newblock arXiv:2303.16455, 2023.

\bibitem{sun2024wanda}
M.~Sun, Z.~Liu, A.~Bair, and J.~Z.~Kolter.
\newblock A simple and effective pruning approach for large language models.
\newblock In {\em Proceedings of ICLR 2024}. arXiv:2306.11695, 2024.

\bibitem{gao2023interpretability}
L.~Gao and L.~Guan.
\newblock Interpretability of machine learning: Recent advances and future prospects.
\newblock arXiv:2305.00537, 2023.

\bibitem{AbboodEtAl2023}
N.~Abbood, J.~Effert, K.~A.~J.~Bozhueyuek, and H.~B.~Bode.
\newblock Guidelines for optimizing type~S nonribosomal peptide synthetases.
\newblock {\em ACS Synthetic Biology}, 12(8):2432--2443, 2023.

\bibitem{niu2023mL-BFGS}
Y.~Niu, Z.~Fabian, S.~Lee, M.~Soltanolkotabi, and S.~Avestimehr.
\newblock {mL-BFGS}: A momentum-based {L-BFGS} for distributed large-scale neural network
optimization.
\newblock arXiv:2307.13744, 2023.

\bibitem{yang2024gas}
Q.~Yang.
\newblock Blending ensemble for classification with genetic-algorithm generated alpha factors
and sentiments ({GAS}).
\newblock arXiv:2411.03035, 2024.

\bibitem{das2023mlwavelet}
N.~Das and M.~Chakraborty.
\newblock Machine learning-driven threshold optimization for wavelet packet denoising in
{EEG}-based mental state classification.
\newblock {\em IEEE Access}, 2023.

\bibitem{Greenside}
P.~Greenside, M.~Hillenmeyer, and A.~Kundaje.
\newblock Prediction of protein-ligand interactions from paired protein sequence motifs and
ligand substructures.
\newblock {\em Pacific Symposium on Biocomputing}, 23:20--31, 2018.

\bibitem{Fu2024}
X.~Fu, Y.~Yuan, H.~Qiu, et~al.
\newblock {AGF-PPIS}: A protein-protein interaction site predictor based on an attention
mechanism and graph convolutional networks.
\newblock {\em Methods}, 222:142--151, 2024.

\bibitem{xiao2024pelpvp}
C.~Xiao, Z.~Zhou, J.~She, et~al.
\newblock {PEL-PVP}: Application of plant vacuolar protein discriminator based on {PEFT}
{ESM-2} and bilayer {LSTM} in an unbalanced dataset.
\newblock {\em International Journal of Biological Macromolecules}, 277:134317, 2024.

\bibitem{sk2023}
\c{S}.~K.~\c{C}orbac{\i}o\u{g}lu and G.~Aksel.
\newblock Receiver operating characteristic curve analysis in diagnostic accuracy studies:
A guide to interpreting the area under the curve value.
\newblock {\em Turkish Journal of Emergency Medicine}, 23(4):195--198, 2023.

\end{thebibliography}

\end{document}